\documentclass[8pt,twocolumn]{extarticle}
\usepackage[utf8]{inputenc}
\usepackage[T1]{fontenc}
\usepackage{amsmath,amssymb,amsfonts}
\usepackage{algorithmic}
\usepackage{graphicx}
\usepackage{graphics}
\usepackage{textcomp}
\usepackage{siunitx}
\usepackage{xcolor}
\usepackage{enumerate}
\usepackage{csquotes}
\usepackage{balance}
\usepackage{array}
\usepackage{natbib}
\usepackage{booktabs}    
\usepackage{listings}
\usepackage{color}
\usepackage{scalefnt}
\usepackage{multirow}
\usepackage{pifont}
\usepackage[framemethod=TikZ]{mdframed}
\usepackage{placeins}
\usepackage{xspace}
\usepackage{url}
\usepackage{balance}
\usepackage{authblk}

\usepackage[most]{tcolorbox}

\definecolor{oldlace}{rgb}{0.99, 0.96, 0.9}
\newmdenv [ %
 skipabove=\topsep,
 skipbelow=\topsep,
 leftmargin       = 2,
 rightmargin      = 2,
 splittopskip     = \topskip]{mh}

\newmdenv [ %
 skipabove=\topsep,
 skipbelow=\topsep,
 roundcorner = 5pt,
 leftmargin = 2,
 rightmargin = 2,
 innertopmargin = 3,
 splittopskip = 3]{mq}

\definecolor{javared}{rgb}{0.6,0,0} 
\definecolor{javagreen}{rgb}{0.25,0.5,0.35} 
\definecolor{javapurple}{rgb}{0.5,0,0.35} 
\definecolor{javadocblue}{rgb}{0.25,0.35,0.75} 
\newcommand{\repo}{\url{https://bit.ly/32ZyBXe}}

\newcommand{\projects}{711\xspace}
\newcommand{\tfs}{1954\xspace}
\newcommand{\tfsu}{192\xspace}

\newcommand{\tfsup}{9.82\%\xspace}

\title{Work Practices and Perceptions from Women Core Developers in OSS Communities}

\author[1]{Edna Dias Canedo}
\author[1]{Rodrigo Bonif\'{a}cio}
\author[1]{M\'{a}rcio Vinicius Okimoto}
\author[2]{Alexander Serebrenik}
\author[3]{Gustavo Pinto}
\author[4]{Eduardo Monteiro}

\affil[1]{Computer Science Department, University of Bras\'{i}lia, Brazil}
\affil[2]{Eindhoven University of Technology, Eindhoven The Netherlands}
\affil[3]{Faculty of Computing, Federal University of Par\'{a}, Brazil}
\affil[4]{Statistics Department, University of Bras\'{i}lia,  Brazil} 

\date{}
\usepackage[strict]{changepage}

\usepackage{framed}

\definecolor{formalshade}{rgb}{0.95,0.95,1}

\newenvironment{formal}{%
  \MakeFramed{\advance\hsize-\width\FrameRestore}%
  \noindent\hspace{-4.55pt}
  \begin{adjustwidth}{}{7pt}%
  \vspace{2pt}\vspace{2pt}%
}
{%
  \vspace{2pt}\end{adjustwidth}\endMakeFramed%
}

\begin{document}

\maketitle

\begin{abstract}
\textbf{Background.} The effect of gender diversity in open source communities has gained increasing attention from practitioners and researchers. For instance, organizations such as the Python Software Foundation and the OpenStack Foundation started actions to increase gender diversity and promote women to top positions in the communities. {\bf Problem.} Although the general underrepresentation of women (a.k.a. horizontal segregation) in open source communities has been explored in a number of research studies, little is known about the \emph{vertical segregation} in open source communities---which occurs when there are fewer women in high level positions. \textbf{Aims.} To address this research gap, in this paper we present the results of a mixed-methods study on gender diversity and work practices of core developers contributing to open-source communities. \textbf{Method.} In the first study, we used mining-software repositories procedures to identify the core developers of \projects open source projects, in order to understand how common are women core developers in open source communities and characterize their work practices. In the second study, we surveyed the women core developers we identified in the first study to collect their perceptions of gender diversity and gender bias they might have observed while contributing to open source systems. \textbf{Results.} Our findings show that open source communities present both horizontal and vertical segregation (only 2.3\% of the core developers are women). Nevertheless, differently from previous studies, most of the women core developers (65.7\%) report never having experienced gender discrimination when contributing to an open source project. Finally, we did not note substantial differences between the work practices among women and men core developers. \textbf{Conclusions.} We reflect on these findings and present some ideas that might increase the participation of women in open source communities. 
\end{abstract}

\section{Introduction}
\label{intro}

Software development often involves the participation and interaction of many contributors, who do not necessarily share the same physical space, culture, and beliefs~\cite{DBLP:conf/chi/VasilescuPRBSDF15}. This diversity might positively influence software development practices and achievements. Previous works reported that gender diversity improves not only teams' productivity, but also the quality of software products~\cite{DBLP:conf/chi/VasilescuPRBSDF15, Catolino:2019:GDW:3339974.3339977, DBLP:conf/group/HuiF16}. 
Even though gender diversity is valued by many software development organizations~\cite{google2019,DBLP:conf/icse/ImtiazMCRBM19}, the field remains dominated by men, and gender bias has been pinpointed as one of the forces that contribute to the underrepresentation of women in the software industry~\cite{DBLP:conf/icse/ImtiazMCRBM19,DBLP:conf/icse/WangWR19}.   

Existing studies report the small number of women contributing to OSS communities, especially in leadership positions~\cite{OpenStack2017,DBLP:journals/software/IzquierdoHSR19,DBLP:conf/msr/RoblesRSVG14,DBLP:conf/msr/0008S16,DBLP:journals/software/IzquierdoHSR19,DBLP:journals/jserd/OrtuDCSTM17,gila2014impact}. \citet{DBLP:journals/software/IzquierdoHSR19} analyzed the percentage of women in positions of governance and leadership in the OpenStack Foundation, reporting an increase in the percentage of women in leadership positions---though the number is still low (around 10--12\%). \citet{nafus2012patches} reported that ``women were sexualized, hurtful and offensive talk was openly defended, and women were obliged to remind men not to stare and point at them''. \citet{DBLP:conf/icse/WangWR19} presented the results of a survey with 142 software engineers in seven OSS organizations and discuss that software engineers regardless their gender implicitly associate software development to a \emph{male activity}. Furthermore, the authors argue that developers express gender biases while taking technical decisions~\cite{DBLP:conf/icse/WangWR19}. Finally, \citet{DBLP:conf/icse/ImtiazMCRBM19} have evaluated presence of several gender biases from the sociological literature in OSS projects.

So far, in the studies of gender and gender bias in OSS no distinction has been made between more and less experienced contributors. Differences between more senior and more junior women have been observed outside the software engineering realm~\cite{10.1371/journal.pone.0225763,doi:10.1177/0956797611417258}. 
We complement the existing literature with the perspective of \emph{vertical gender segregation}~\cite{benschop2006small,campos2011patterns}, which deals with distribution inequalities within organization levels. Note that OSS communities are concerned with both horizontal (i.e., the general underrepresentation of women in OSS) and vertical gender (i.e., the participation of women in high level positions) diversity. The Python community, for instance, has started an effort to increase diversity in its core development team\footnote{http://pyfound.blogspot.com/2019/02/the-north-star-of-pycascades-core.html} and OpenStack Foundation increased the percentage of women in 	the Technical Committee from 0\% to 15\%~\cite{OpenStack2017}. 

The goal of this paper is twofold: first, we explore the issue of vertical segregation in open source communities and, second, we study the work practices and perceptions of gender bias--- from the point of view of \emph{women core developers} that contribute to OSS projects. Altogether, we answer the following research questions: 

\begin{enumerate}[(RQ.1)]
\item How common are women core developers in OSS?  
\item Are there differences in the work practices of women or men core developers?\footnote{We compare women with \emph{men} since OSS is known to be male-dominated~\cite{DBLP:conf/msr/RoblesRSVG14}. Study of development practices of non-binary software developers should be a topic of a separate study.} 
\item How do women core developers perceive gender diversity and gender bias in OSS communities? 
\item What are the actions women core developers consider important to make OSS communities more inclusive?  
\end{enumerate}

In the first two research questions we address the issues of vertical segregation and work practices of core developers. In the third and fourth research questions we address the perceptions of women core developers on gender bias. To answer these questions, we carry out a mixed-method study. We first identify the core developers of open-source systems, by mining the source code history of more than 700 OSS projects. Core developers here are those developers that significantly contribute to the development of a system; and thus the continuity of a project might be compromised in the case they decide not to contribute to the development of a system anymore. We identify core developers using \emph{Truck Factor}~\cite{DBLP:conf/profes/RiccaMT11, DBLP:conf/wcre/CosentinoIC15, DBLP:conf/iwpc/AvelinoPHV16}.
To identify the gender of \emph{core developers}, we leverage two gender classification algorithms: GenderComputer \cite{DBLP:journals/iwc/VasilescuCS14} and Namsor \footnote{https://www.namsor.com/}. 

Considering the \emph{intersection} of the results of both gender classification tools, we found 42 women core developers, and we were able to manually confirm the gender of 36. We invited them to answer a survey about gender bias on OSS communities---getting answers from 35 of them (97.22\% of response rate). Our study produced a set of findings; we highlight three of them next:

\begin{enumerate}
  \item While 5.35\% of all contributors are women, the percentage of women among core developers drops to 2.30\% (characterizing both horizontal and vertical gender segregation.)
 \item There is no significant difference between the \emph{work practices} between women and men core developers.  
  \item Gender bias also occurs among core developers. 34.3\% of women core developers surveyed state having observed gender bias at least once while contributing to OSS projects.
\end{enumerate}
\section{Background and Related Work}
\label{related-work}

Gender diversity in the fields of Science, Technology, Engineering, and Maths (STEM) has been investigated by multiple researchers and gained considerable public attention in recent years. Several educational institutions design programs with the aim of reducing the gender gap among students, which, in a long term, might lead to a positive impact on the gender diversity of teams in the workforce~\cite{DBLP:conf/icse/Borsotti18, DBLP:journals/ijhcitp/BhattacharyaBM18, DBLP:conf/icse/MacielBF18,DBLP:journals/jucs/BotturiBM12, DBLP:conf/sigcse/RheingansDDI18,DBLP:conf/icse/RibaupierreJLC18,DBLP:conf/sigcse/LopezSG05, DBLP:journals/entropy/BotellaRLM19}. 
Although there are some conflicting findings about the effects of diversity on team performance~\cite{DBLP:conf/chi/VasilescuPRBSDF15,gila2014impact,DBLP:conf/group/HuiF16}, some reports show evidence that promoting gender-balanced teams improves innovation and problem-solving capacity, as well as leads to a healthier work environment~\cite{DBLP:conf/group/HuiF16}. 

An inclusive work environment should mitigate possible conflicts that might arise due to diversity. Several authors study relation between gender diversity and performance of software development teams~\cite{gila2014impact,DBLP:conf/chi/VasilescuPRBSDF15,DBLP:journals/jserd/OrtuDCSTM17,DBLP:conf/esem/BosuS19}. 
Hui and Farnham \cite{DBLP:conf/group/HuiF16} seek to understand how interpersonal practices and the use of socio-technical tools can promote gender diversity and help to form more independent innovative teams. Catolino et al.~\cite{Catolino:2019:GDW:3339974.3339977,DBLP:journals/software/CatolinoPTSF20} have studied the relation between gender diversity and ineffective communication. 
 
Another line of research has focused on retention of women in OSS.
Qiu et al. \cite{DBLP:conf/icse/QiuNBSV19} have shown that involvement in teams using diverse technologies is beneficial for duration of engagement of women in OSS. Balali et al.~\cite{DBLP:journals/cscw/BalaliSASG18} argued that duration of engagement of women in OSS is negatively affected by differences in the viewpoint of men and women mentors about gender personalities;  underestimation of women’s capabilities by both open source community and women newcomers themselves; and ignorance of men mentors’ about the community being harsh to women.

Yet another group of studies have focused on gender biases in software development~\cite{DBLP:conf/esem/BosuS19,DBLP:conf/icse/WangWR19,DBLP:conf/icse/ImtiazMCRBM19}. For example, Imtiaz et al.~\cite{DBLP:conf/icse/ImtiazMCRBM19}  concluded that while the effects of gender bias are virtually invisible on the investigated projects, women restrict their involvement to fewer projects and organizations, in comparison with men developers. 

Lee and Carver~\cite{DBLP:conf/icse/LeeC19} carried out an investigation of the men and women perspectives on gender relations in Free/Libre OSS projects. The study found, that while some respondents expressed a positive feeling about women's participation, some contributors were strongly opposed to their inclusion. Women reported the difficulty of being accepted in the community and the gender-biased comments of colleagues as major barriers for their participation.

In summary, the existing research on gender diversity in software engineering seeks to investigate the benefits of diversity on software teams and to try to understand the possible causes of the underrepresentation of women. 
Unlike the previous work, we investigate the vertical segregation problem in a comprehensive number of open source communities. We also study the work practices as well as perceptions of women core developers on gender bias. Finally, we survey the women core developers to identify actions that should be taken to make OSS more inclusive.

\section{Study Settings}\label{sec:methodology}

The main goal of this research is to improve our understanding on work practices and gender bias in open source communities, focusing on a particular group of contributors: \emph{women core developers}. To achieve this goal, we use a mixed-methods approach. First we mine open source repositories to identify women core developers and to understand their work practices when contributing to open source communities. Second, we conduct a survey with women core developers, to understand their perceptions about gender bias in open source communities. 

\subsection{Settings for the first study: Mining open source repositories}

 Our approach for mining open source repositories has five steps. {\bf In the first step} we used \emph{purposeful sampling}~\cite{baltes2020sampling} to build a dataset of open source projects from different domains and written in different programming languages. To this end we use the GitHub API to search for the 100 most popular projects written in the 15 most popular programming languages at GitHub.
 To operationalize popularity of programming languages we use a recent report\footnote{\url{https://octoverse.github.com/2018/projects\#languages}}, of projects---the number of stars~\cite{casual-developers,gh-popularity}. This dataset comprises open source projects of different sizes, targeting different domains (from compilers to mobile apps), and written in a diversity of languages, e.g., scripting languages such as Shell Script, system programming languages such as C and Go, and languages often used for web- and mobile development such as TypeScript and Swift.  

As we study core developers, we focus on  ``sufficiently large'' projects with ``sufficiently many'' committers.
To determine the thresholds we compute the first quartiles of the distribution of SLOC and number of committers, and exclude projects having less SLOC or less committers than the thresholds.
In this way we preserve \projects projects written in 14 languages\footnote{No Objective-C projects meet the thresholds.} with at least \num{5183} SLOC and \num{33} committers. Tables~\ref{tab:projects-statistics} and~\ref{tab:freqall} present descriptive statistics. 

{\bf In the second step}, we identify the core developers.
To this end we use  the notion of the Truck Factor (TF). 
``TF developers'' is the minimal set of developers a project depends on for its maintenance and evolution, i.e., if the ``TF developers'' abandon the project (e.g., after being hit by a truck) the project maintenance will be heavily affected.
We call ``TF developers'' \emph{core developers}.
Indeed, Ricca et al.~\cite{DBLP:conf/profes/RiccaMT11} state that the TF can be used to assess the distribution of project knowledge among developers; and 
Bosu and Sultana argue that TF is a proxy for identifying ``\emph{developers that made significant contributions to guide the development and evolution of the project}''~\cite{DBLP:conf/esem/BosuS19}. Several approaches to compute the TF have been proposed in the literature. In our paper, we use the approach of Avelino et al.~\cite{DBLP:conf/iwpc/AvelinoPHV16}, shown to outperform competing approaches ~\cite{DBLP:conf/iwpc/FerreiraVF17}. We have identified \tfs core developers in \projects projects. 

{\bf In the third step} we identified
the gender of all core developers using two gender identification tools, GenderComputer \cite{DBLP:journals/iwc/VasilescuCS14} and Namsor\footnote{https://www.namsor.com/}. 
In the case of a disagreement between the tools, we assign ``Unknown'' to a given core developer. 
From \tfs core developers, the tools disagree in \tfsu cases (\tfsup). 
After identifying the gender of the core developers, {\bf in the fourth step} we proceeded to collect the contributions from women core developers (WCD). To compare contributions of women with those of men, we randomly select three samples of male core developers (MCD1, MCD2, MCD3)---with the same number of members as WCD. We used the GitHub API to collect all commits and pull requests (PRs) from the contributors in WCD, MCD1, MCD2, and MCD3. We classify the size and the type of contributions using an approach of Hattori and Lanza~\cite{hattori:ase-2008}. 

Finally, {\bf in the last step (data analysis)} we  used Exploratory Data Analysis (EDA)~\cite{data-analysis-using-R} to (a) characterize how common are women core developers in open source projects and (b) understand how women core developers contribute to open source projects. EDA covers different statistics (e.g., median and mean) and graphical methods (e.g., histograms and boxplots) to build a general understanding about  the data distribution~\cite{data-analysis-using-R}. In addition, we leveraged a  statistical procedure (\texttt{nparcomp}) of Konietschke et al. ~\cite{JSSv064i09} for performing a multiple comparison on the work practices of the experimental groups (MCD1, MCD2, and MCD3) against the control group (WCD).  Similarly to a previous work
~\cite{vasilescu:social2013}, we set the \texttt{nparcomp} analysis to use the Dunnett-type contrasts~\cite{dunnett1955multiple} and the \emph{probit} transformation function (as the asymptotic approximation method). All datasets and scripts are available online (\repo).
\begin{table*}[ht]
 \centering
 \begin{tabular}{rrrrrrr}
   \hline
             & Min. & 1st Qu. & Median & Mean & 3rd Qu. & Max. \\ \hline
 Lines of Code           & \num{5191}  & \num{19523} & \num{57013} & \num{259367.63} & \num{195265} & \num{9442645} \\
 Num. of Contributors    & \num{33}   & \num{80} & \num{145} & \num{292.77} & \num{297} & \num{8413} \\
 Num. of Forks           & \num{54}   & \num{774} & \num{1481} & \num{2949.94} & \num{3171} & \num{64712} \\
 Num. of Watchers        & \num{1145} & \num{5882} & \num{9039} & \num{14284.96} & \num{16418} & \num{300666} \\
    \hline
 \end{tabular}
 \caption{Descriptive statistics about the projects used in the study}
 \label{tab:projects-statistics} 
 \end{table*}

\subsection{Settings for the second study: Survey with women core Developers} 

To answer the research questions RQ.3 and RQ.4, we conducted a survey with the women core developers we identified in the first study. 
We follow the recommendations of Kitchenham and Pfleeger~\cite{DBLP:books/sp/08/KitchenhamP08}, and organized our survey in six steps. 
Regarding {\bf the first step} (planning), our goal is to capture beliefs of women contributors identified as core developers about (a) gender bias in OSS communities and (b) possible ways for OSS communities to become more inclusive with respect to gender. 

\begin{figure*}[htb]
\includegraphics[scale=0.5]{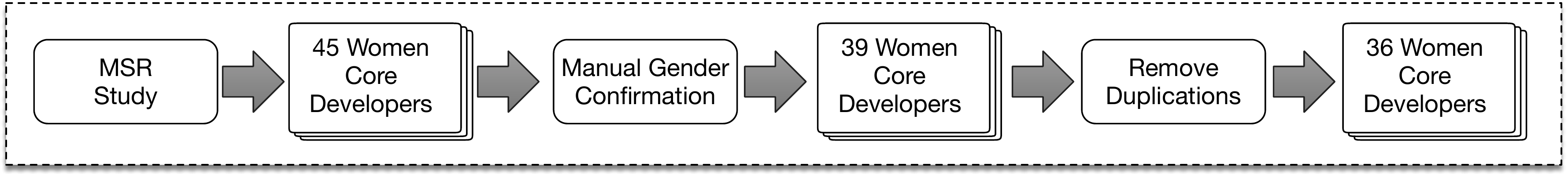}
\caption{Procedure used to define the target population of the survey}
\label{fig:workflow}
\end{figure*}

In {\bf the second step} (defining the target population), we \emph{manually confirm} the gender of the women core developers 
we identified in the first study, using information from social networks (e.g., Facebook, Google+, and Twitter). This activity 
was necessary because our goal was to only gather information from women core developers, and thus we followed a conservative 
approach before inviting our target population to answer our survey. Accordingly, our target population comprises 
36 (distinct) women core developers, which we manually confirm the gender. We could not confirm the gender of 6  
women core developers (out of 45). 
Moreover, among the 39 women core developers, three of them are core developers contributing to two projects in our dataset. Figure~\ref{fig:workflow} shows the procedures we followed to define our target population. 

In {\bf the third and fourth steps} we designed an online survey and validated it using an iterative approach (the first, third, and fourth authors were responsible for reviewing and validating the questions). 
The final version of the survey contains 18  questions (14 closed questions using a Likert scale and 4 open-ended questions), organized in three sections: demographics, contribution to open source communities, and perceptions about gender bias in open source communities. In the \emph{demographics section}, the survey covers information such as age and academic degree of the participants. 
In the \emph{contribution to open source communities} section, the survey presents questions to characterize the engagement of the participants into open source communities, including questions such as \emph{How long have you been contributing to OSS communities?}, \emph{How often do you interact with other team members in OSS communities?}, and \emph{Are you happy with your participation in OSS communities?}. All the questions are available in the paper website (\repo). In the third section (\emph{perceptions of gender-bias in open source communities}) we look at gender bias in OSS communities, including questions such as \emph{Have you ever felt that one of your contributions was not well received due to your gender?}, \emph{How often do you feel that your contributions were not well received due to your gender?} \emph{What would you recommend to increase women's participation in  OSS communities?}. 

In {\bf the fifth step} (conducting the survey) we contacted (via email) the 36 women core developers that correspond to our target population. Over a period of three weeks, we received answers from 35 women core developers (a response rate of \num{97.22}\%). In the email message, we sent the goal of our research and a link to the online survey. Most of the participants agreed to answer the survey without any additional clarification. For instance, one participant answered ``\emph{I was really glad to participate, thank you!''} and other perceived the relevance of the research, answering ``\emph{Thanks for taking up this issue!}''. Nonetheless, other participants requested us to provide additional details before answering the survey, such as our credentials and affiliation. Finally, in the {\bf sixth step} (analysing the results), we leverage exploratory data analysis to consolidate the answers to the Likert scale based questions (in terms of descriptive statistics and plots) while 
the answers to the survey's open-ended questions were literally quoted (all datasets are available in the paper website \repo). 

\section{Results of the Fist Study: Mining Software Repositories}
\label{resultfirst}

\begin{figure}[h!]
  \centering
  \includegraphics[width=0.8\linewidth, clip = true, trim= 0px 40px 40px 60px]{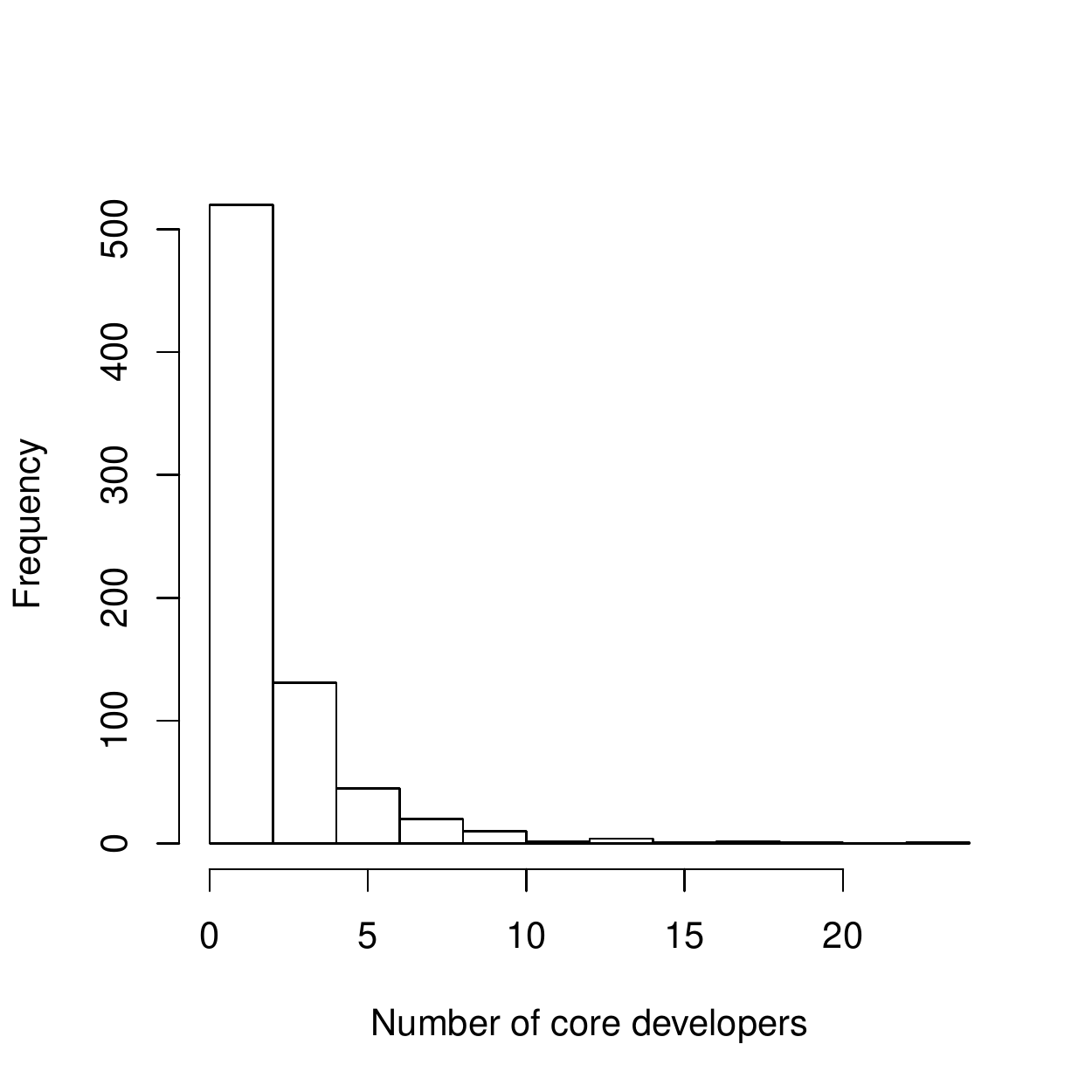}
  \caption{Histogram with the number of core developers}
  \label{fig:histogram}
\end{figure}

We first report the results of an exploratory data analysis. Figure~\ref{fig:histogram} shows a histogram with the number of core developers in each project: most of the projects have only one core developer.  
This finding corroborates the work of Avelino et al.~\cite{DBLP:conf/esem/AvelinoCVS19}, which reports that most projects have a small number of TF developers and that the TF algorithm reveals just one core developer in 57\% of the projects. This situation might represent a risk, since by definition of the Truck Factor~\cite{DBLP:conf/iwpc/AvelinoPHV16}, the continuity of a project might be compromised when a single core developer decides to leave it.

Nonetheless, we found projects having more than 5 core developers, including \textsc{elasticsearch} (17 core developers) and the implementation of Python and Go programming languages (15 and 9 core developers, respectively). 
Considering all \projects projects, 88 projects (12.37\%) have at least five core developers; and we found a small correlation between lines of code and the number of core developers of a project (Spearman's $\rho = 0.30$ with p-value $= 0.001$). 
Also, there is a moderate correlation between the number of contributors and the number of core developers of a project ($\rho = 0.41$ with p-value $< 0.01$). Our dataset comprises \num{1954} core developers, from which 235 developers are core developers in more than one project. 

\subsection{How common are women core developers in OSS communities?}
45 core developers are recognized both by  GenderComputer and Namsor as women (2.30\%), while 1,717 are recognised as men (87.87\%). We could not confirm the gender of 192 core developers (9.82\%) due to disagreement between the tools. 
We found women core developers in 37 (5.24\%) out of the \projects GitHub projects considered in our analysis. 
We also found women core developers in projects written in all programming languages we consider in our study. Interestingly seven projects written in TypeScript (10.93\%) have at least one women as core developer; though only 2.17\% of projects using PHP have at least one women as core developer. Table~\ref{tab:freqall} summarizes these findings. 
 
The barplot of Figure~\ref{fig:bp-language} presents a different perspective: the percentage of women core developers (over the total of core developers) considering the different programming languages. 
That is in our dataset less than one percent of core developers in projects using PHP and Shell programming languages are women. Contrasting, more than four percent of the core developers in projects written in Swift and TypeScript are women. Altogether, we answer our first research question (``How common are women key developers in OSS projects?'') as follows:

\begin{figure}[h!]
\centering{
\includegraphics[width=0.8\linewidth]{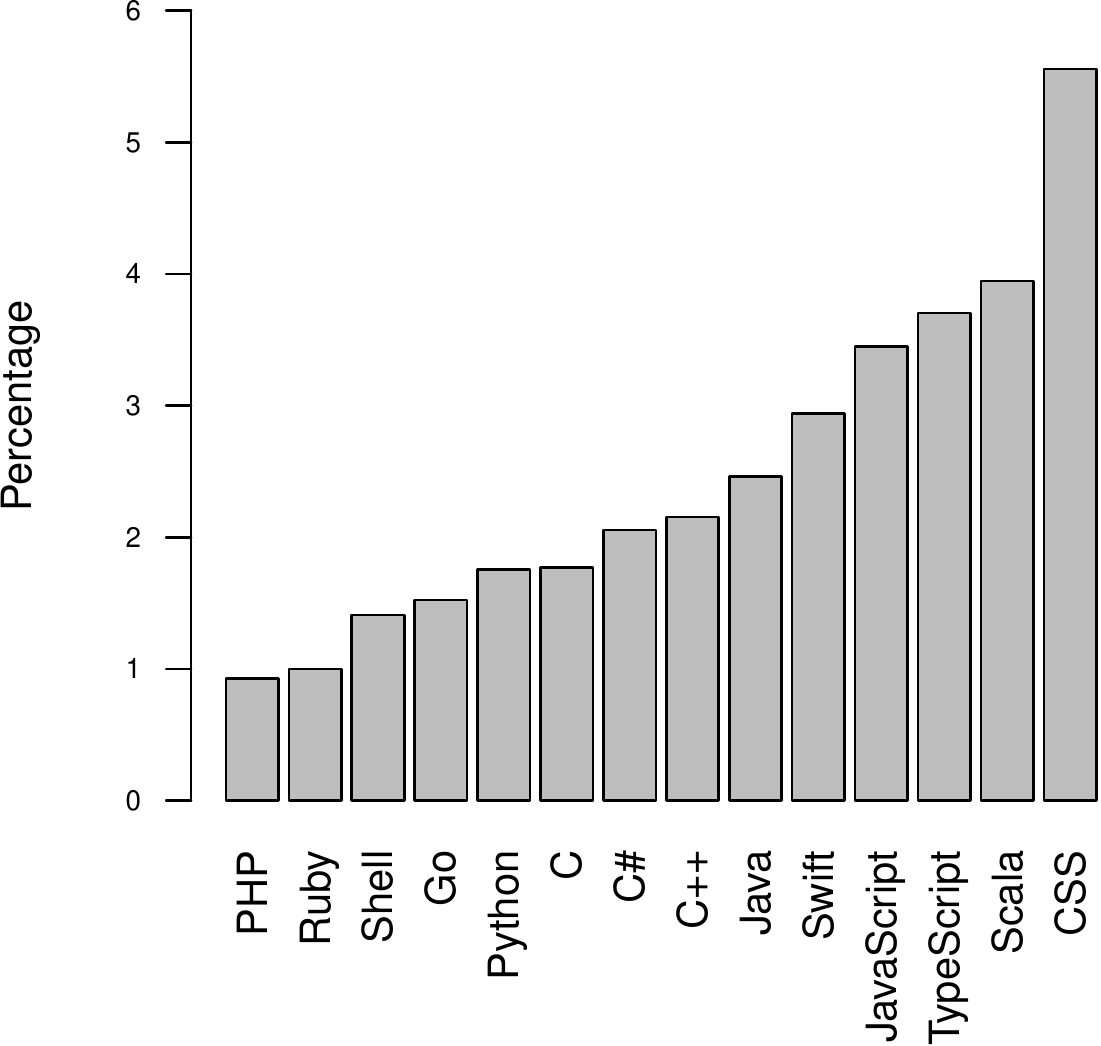}}
\caption{Percentage of women core developers in projects written in different programming languages}
\label{fig:bp-language}
\end{figure}

\begin{mq}
\emph{Among \projects GitHub projects, we identified 1954 core developers. 45 core developers (2.30\%) are identified as women. Since the percentage of developers identified as women in our dataset is 5.35\%, these findings suggest an underrepresentation of women core developers in OSS projects, i.e., vertical gender segregation. }
\end{mq}

Besides gender bias, other factors (such as the reward model and the possible long term benefits of contributing to open source projects) could contribute to this  underrepresentation of women core developers in OSS projects. 
Considering this quantitative assessment, we can mostly report on the extent of this underrepresentation---and thus we postpone a discussion of possible causes for this underrepresentation (and how to deal with them) to the next section. 

\subsection{Are there differences in the work practices of women and men core developers?}

\begin{figure}[h!]
\centering{
\includegraphics[width=0.90\linewidth]{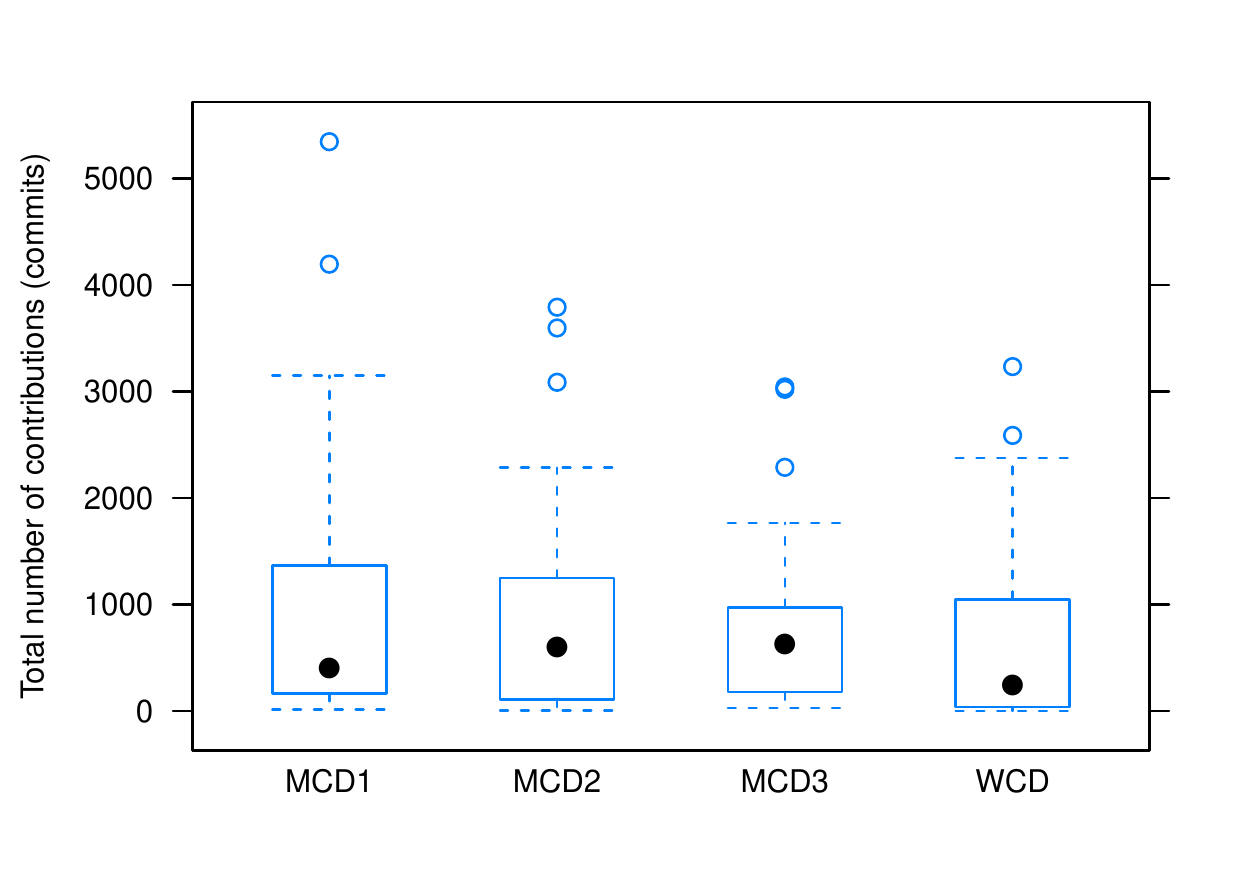}}
\caption{Number of commits from the four sets of core developers. WCD stands for Women Core Developers while MCD* stands for the random sets of Men Core Developers}
\label{fig:bp-commit}
\end{figure}

To better understand the work practices of women core  developers, we also explore two additional questions: (a) How do the number, frequency, and size of contributions of women core developers compare to the number, frequency, and size of contributions of men core developers? and (b) How do the \emph{types of contributions} of women core developers differ from the \emph{types of contributions} of men core developers? 
Accordingly, we mined the commit history from the 36 women core  developers (WCD) whose gender we could manually confirm, out of the initial set of 45 we identified (see Figure~\ref{fig:workflow}). To counterbalance the effects of randomness in selecting samples, we randomly generated three datasets with men core developers (MCD1, MCD2, and MCD3), and collected their commit history. Each one of these datasets comprise 36 men core developers. 
We contrast the working practices of the experimental groups (MCD1, MCD2, and MCD3) with our control group (WCD), using the \texttt{nparcomp} procedure (see Section~\ref{sec:methodology}).
Figure~\ref{fig:bp-commit} shows the total number of contributions (commits) from these sets (MCD1, MCD2, MCD3, and WCD). Figure~\ref{fig:bp-commit} might suggest a small difference in terms of the distribution of the total number of commits. Nonetheless, the results of the multiple comparison \texttt{nparcomp} test do not reveal any significant difference at 5\% level. 
To measure the frequency of the commits, we use three auxiliary metrics: 
\emph{Max Date}, \emph{Min Date}, and \emph{Distinct Dates}. \emph{Max Date}  (\emph{Min Date}) corresponds to the date of the last (first) commit of a core developer, in one of the sets MCD1, MCD2, MCD3 or WCD. \emph{Distinct Dates} corresponds to the number of distinct commit dates of a developer, again, in one of the sets MCD1, MCD2, MCD3 or WCD. Finally, we compute the \emph{Frequency} of commits using Eq. (1).

\begin{table*}[ht]
\caption{Summary of the dataset with core developers. MCD means the number of men core developers, WCD means the number of women core developers, and UCD means the number of core developers we could not identify the gender. PWCD means the number of projects with women core developers, and PPWCD corresponds to the percentage of projects with women core developers.}
\centering
\begin{small}
\begin{tabular}{rlrrrrrrr}
 \toprule
  & Language & Number of Projects & Number of Contributors & MCD & WCD & UCD & PWCD & PPWCD  \\ 
   \midrule
 1 & C &  50 & \num{11627} &   99 &   2 &  12 &   2 & 4.00 \\
 2 & C\# &  63 & \num{10150}  & 129 &   3 &  14 &   3 & 4.76 \\
 3 & C++ &  67 & \num{16954} & 191 &   5 &  36 &   3 & 4.48 \\
 4 & CSS &  23 & \num{2505}  &  30 &   2 &   4 &   2 & 8.70 \\
 5 & Go &  68 & \num{19232}  & 169 &   3 &  25 &   3 & 4.41 \\
 6 & Java &  44 & \num{9575}  & 109 &   3 &  10 &   2 & 4.55 \\
 7 & JavaScript &  67 & \num{33899}  & 172 &   7 &  24 &   5 & 7.46 \\
 8 & PHP &  46 & \num{13315}  &  98 &   1 &   9 &   1 & 2.17 \\
 9 & Python &  42 & \num{20706}  & 155 &   3 &  13 &   2 & 4.76 \\
10 & Ruby &  60 & \num{36064}  & 183 &   2 &  15 &   2 & 3.33 \\
11 & Scala &  57 & \num{8956}  & 137 &   6 &   9 &   5 & 8.77 \\
12 & Shell &  24 & \num{5417}  &  63 &   1 &   7 &   1 & 4.17 \\
13 & Swift &  36 & \num{4129}  &  59 &   2 &   7 &   2 & 5.56 \\
14 & TypeScript &  64 & \num{20589} & 123 &   5 &   7 &   4 & 6.25 \\
    \bottomrule
    \multicolumn{2}{r}{Total} & 711 & \num{213118} & \num{1717} & 45 & 192 & 37 &  mean = 5.24, sd = 1.92
 \end{tabular}
 \end{small}
 \label{tab:freqall}
 \end{table*}

\begin{small}
\begin{eqnarray}
\mbox{\it Frequency} & = & \frac{\mbox{\it Distinct\ Days}}{\mbox{\it interval}(\mbox{\it Min\ Date}, \mbox{\it Max\ Date})} \times 100
\end{eqnarray}
\end{small}

\begin{figure}[h!]
\centering{
\includegraphics[width=0.92\linewidth]{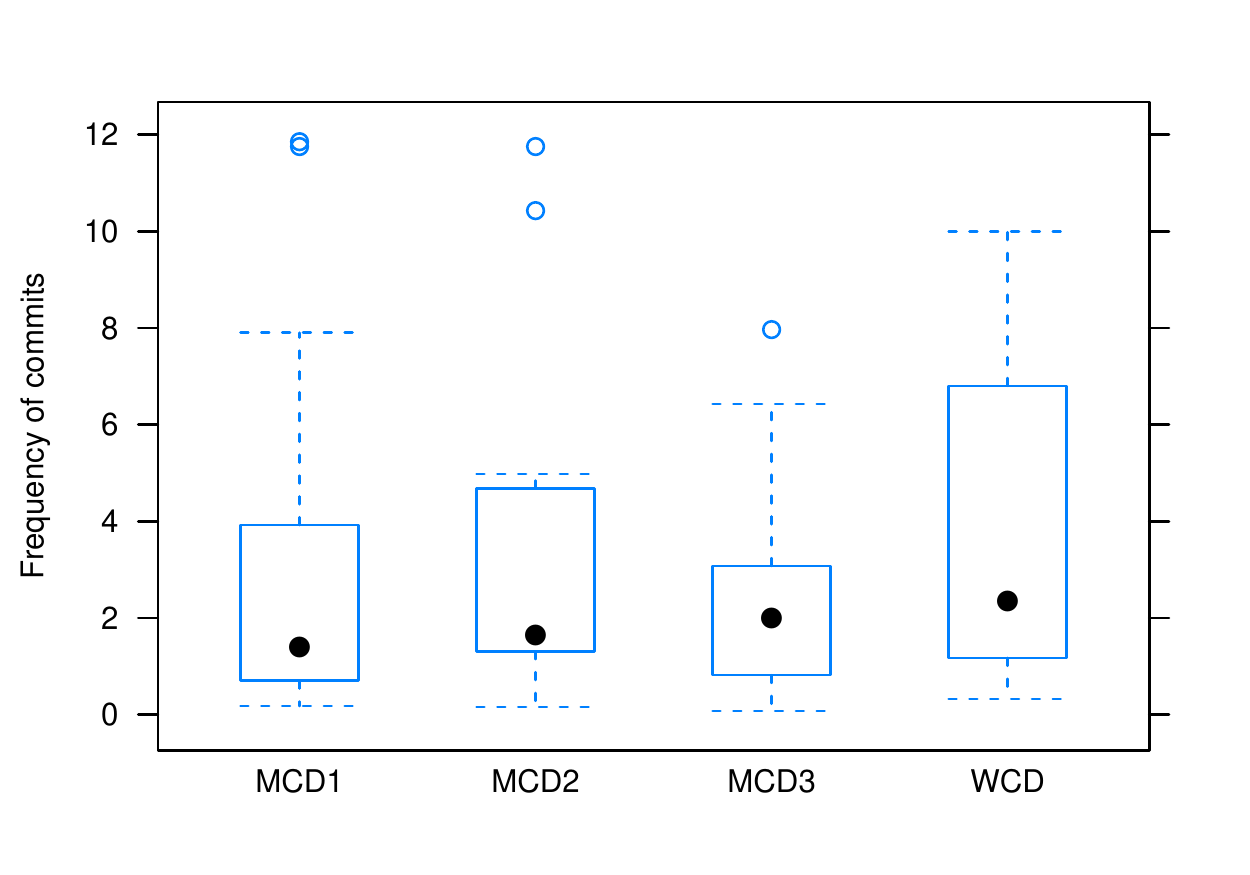}}
\caption{The frequency of commits from core developers. This frequency corresponds to the percentage of days the core developers contribute to a project, considering an interval from the first and last commits to the project.}
\label{fig:bW-frequency}
\end{figure}

\begin{figure*}[t!]
\centering{
\includegraphics[width=0.9\textwidth, clip = true, trim= 0px 20px 10px 20px]{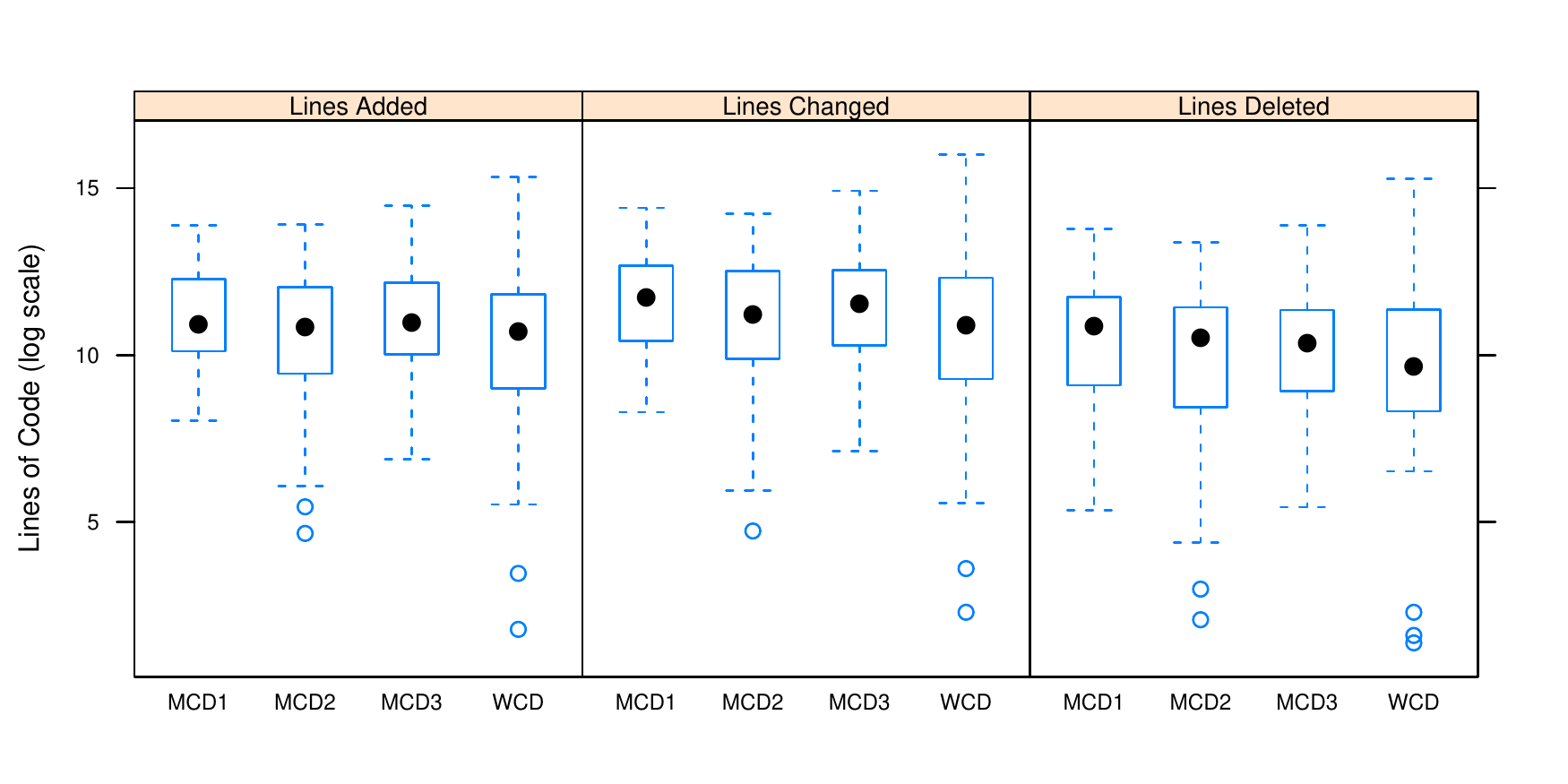}}
\caption{Log-scale of the size of the contributions, in terms of lines of code added, changed, and deleted}
\label{fig:bp-size}
\end{figure*}

\noindent The boxplots of Figure~\ref{fig:bW-frequency} summarizes the frequency of commits per group (WCD, MCD1, MCD2, and MCD3). 
Descriptive statistics suggest that WCD commit code more frequently than MCD (the median value of the frequency of commits in WCD is 2.34\%, while the median value of the frequency of commits in MCD1 is 1.38\%, in MCD2 is 1.64\%, and in MCD3 is 2.01\%). However, these differences are not statistically significant (all $p$-values reported by the \texttt{nparcomp} test exceed 0.3).
Regarding the size of the contributions, we computed the total lines of code (and the number of files) \emph{added}, \emph{changed}, and \emph{deleted}, from the set of contributions of MCD1, MCD2, MCD3, and WCD. Figure~\ref{fig:bp-size} summarizes the (log-scale) size of contributions in terms of lines of code. Again, no statistically significant differences could be observed. That is, considering these results, we conclude that there is no difference in terms of the number, frequency, and size of commits with regards the gender of core developers.

\begin{mq}
\emph{We could not find statistically significant differences in the practices of women and men w.r.t. the number, frequency, and size of commits.}
\end{mq}

Finally, we used the approach of Hattori and Lanza~\cite{hattori:ase-2008} to investigate the differences on the type of contributions from MCD1, MCD2, MCD3, and WCD. Their approach  
classifies the contributions 
by searching for a set of keywords in the commit message---assigning a commit to a class whenever it finds the first keyword in the commit message. A commit could be classified as \emph{forward engineering}, \emph{reengineering}, \emph{corrective engineering}, 
and \emph{management}. 
Regardless of its simplicity, the assessment of this algorithm has shown a good performance (F-measure = 0.70)~\cite{hattori:ase-2008}. Our dataset comprises a total of \num{115922} commits (\num{22326} from WCD and \num{33840}, \num{32818}, and \num{26938} from MCD1, MCD2, 
and MCD3, respectively). To avoid unbalanced problems 
in this analysis, we \emph{undersample} the set of commits in MCD1, MCD2, and MCD3, and thus we only consider \num{22326} commits coming from contributors in each set. 

We found a statistically significant difference between the size of 
description of the commits (in terms of number of characters) from women core developers and the size of commits' descriptions from men core developers (with a p-value < 0.0001). 
That is, based on our dataset of commits we consider here, women core developers tend to present a more detailed message explaining their contribution changes. This might indicate the gender bias symptom named \emph{proving-it-again}~\cite{DBLP:conf/icse/ImtiazMCRBM19}, which occurs when a group of people that does not align to the default stereotypes has to demonstrate more evidence about their competence.
Since we classify the type of a contribution considering the commit message, this result impacts the number of commits that we could not classify using the Hattori and Lanza method, which is lower when we consider the WCD set. 

\begin{figure*}[h!]
\centering{
\includegraphics[width=\textwidth]{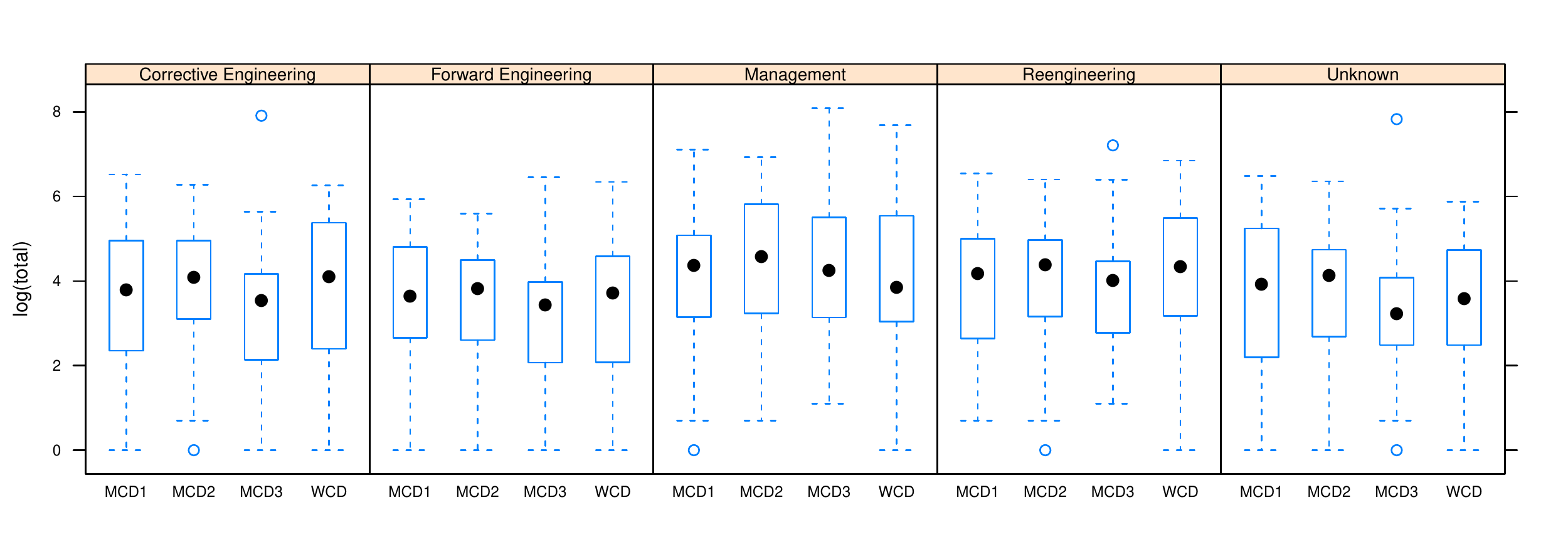}}
\caption{Log-scale of the classes of contributions, according to the approach of Hattori and Lanza~\cite{hattori:ase-2008}}
\label{fig:bp-contribution}
\end{figure*}

Figure~\ref{fig:bp-contribution} 
summarizes the class of contributions using a log scale. The boxplots suggest differences in the types of contributions 
when we consider the different groups. For instance, it seems that the set of WCD contributes more with \emph{Corrective} and \emph{Reengineering} activities; while the sets of men core  developers contribute more with \emph{Management} activities and activities that we were not able to classify (\emph{Unknown}---according to the Hattori and Lanza method). There is no much difference in the \emph{Forward Engineering} activities. We actually found that these differences are statistically significant (with a p-value < 0.0001) using the Chi-squared test. This test is useful to investigate if two categorical variables (gender and type of contributions) have a significant correlation. 

\begin{mq}
\emph{We found statistically significant gender-related differences in the kinds of contributions: women core developers tend to contribute more with reengineering tasks.}
\end{mq}

\section{Results of the Second Study: A Survey with Women Core Developers}
\label{resultsecond}

When analyzing the answers to our survey, we found that women core developers are in general young: 51.4\% of the respondents are between 18 and 25 years old and almost 80\% of the respondents are younger than 35 years old. Regarding academic degree, most of them are undergraduate students (31.4\%); 34.3\% hold a bachelor degree, 11.4\%---a master degree, and 22.9\%---PhD. 77.2\% contribute to open source projects for less than five years: more then twice than percentage of women contribute to open source projects for less than five years reported in the FLOSS 2013 survey~\cite{DBLP:conf/msr/RoblesRSVG14}.

\begin{figure}[h!]
    \begin{centering}
\[\arraycolsep=8pt
    \begin{array}{cc}
         (a) & (b) \\
         {\includegraphics[width=0.45\linewidth]{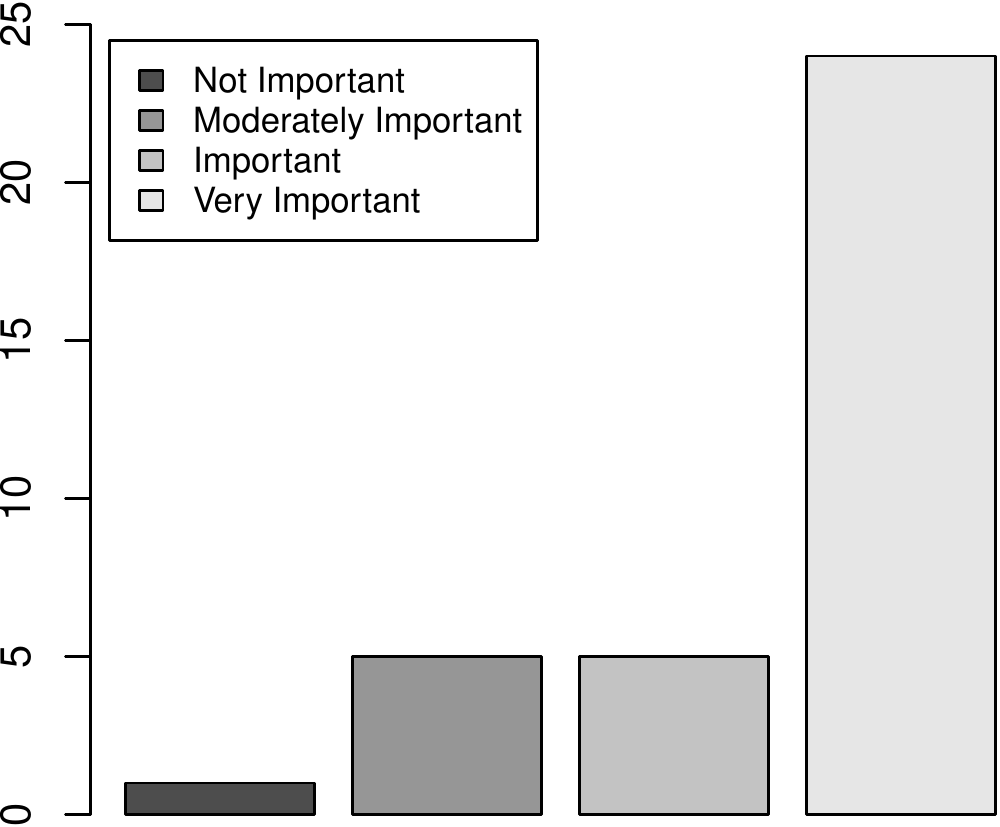}} &  {\includegraphics[width=0.45\linewidth]{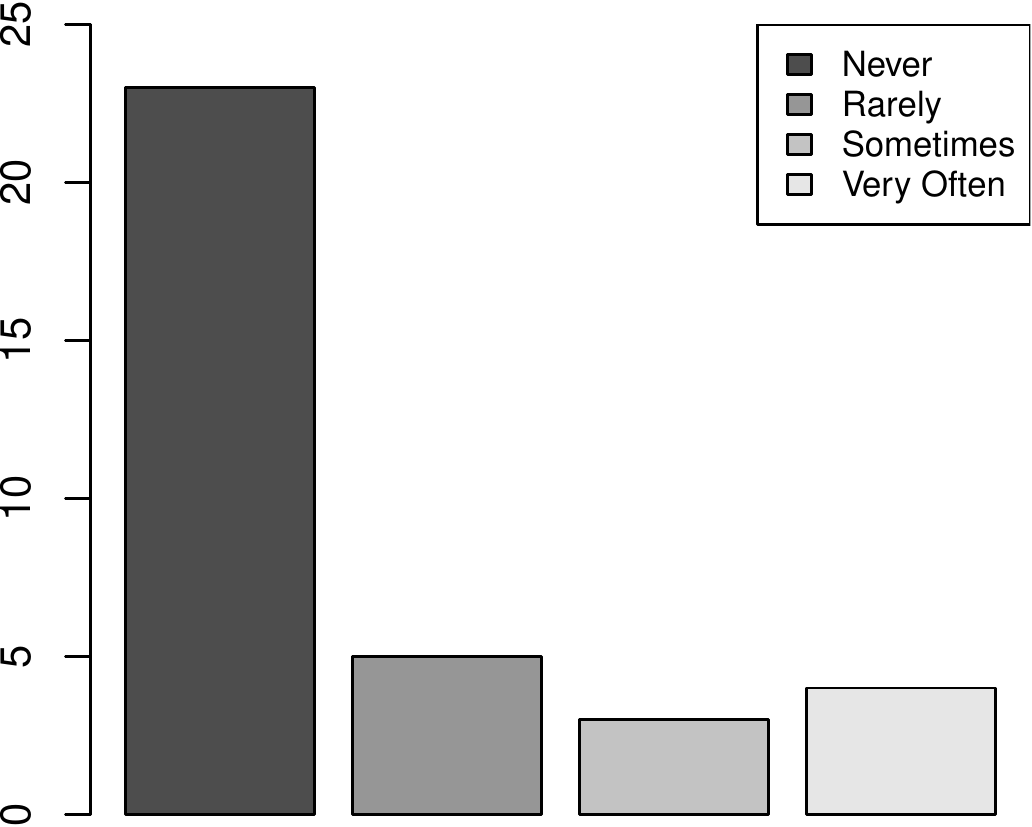}} \\
    \end{array}
    \]
    \end{centering}
    \caption{Figure (a) shows the importance of gender diversity in OSS Communities, while (b) shows how often the respondents feel that their contributions were not well received due to their gender.}
    \label{fig:question11-12}
\end{figure}

Figure~\ref{fig:question11-12}-(a) summarizes perceived importance of gender diversity in OSS communities. 68.6\% of the women core developers consider gender diversity in OSS communities to be very important. According to the respondents, gender diversity can improve team communication, and attract new contributors: these opinions concur with the findings of Catolino et al.~\cite{DBLP:journals/software/CatolinoPTSF20}. For instance, one of the participants states that:

\begin{formal}
\emph{\ldots the interaction between team members, and communication both within the team, as well as with the larger community has to be open and consistent. I believe having more gender diversity can help in this direction, as women may bring a new perspective, and focus more on communication and human aspects. This will help with both building a stronger core team that stays with the project, as well as attracting new contributors. If the team is more welcoming to new members of any gender, then there is a larger pool of potential contributors, and a better chance of them wanting to get involved.} 
\end{formal}

Other respondent states that diversity can contribute to the design of products based on a broader variety of past experiences, promote empathy and build a safer community. 

\begin{formal}
\emph{I cannot say what women can bring exactly, but what I know is that everyone has different experiences in life, as women, men, people of color, members of LGBTQ+ communities, people with disabilities, and I am sure that we need people with different experiences to build products, whether in closed source or in open source. Also because in open source you can get anyone in the world to contribute to a project, different people have different sensibility to different ways of working and communicating. Having people from different cultures and walks of life brings more empathy and therefore potentially a safer environment for people to contribute.}
\end{formal}

Yet another respondent states that gender diversity might not directly improve the productivity of the teams. However, it might increase the design space when conceiving a product's features, which could also be more generic and inclusive.   

\begin{formal}
\emph{I have no idea whether diversity improves productivity.  However, I have found that having diverse development pools (meant in the widest sense) ensures that the developed software is more fit for purpose and generalizable as the diverse experiences ensure that people are thinking about the design from different angles.}
\end{formal}

Figure~\ref{fig:question11-12}-(b) summarises the answers to the question ``\emph{How often do you feel that your contributions were not well received due to your gender?}'' 
Even in a population of core developers, one third of the participants believe that, at least one of their contributions had not been accepted due to gender bias. Moreover, 11.4\% (very) often recognize gender bias while someone appraises their contributions. According to one respondent (P19) (see TableSM1 in \repo), gender bias appears whenever a contribution from a women developer receive less positive feedback: ``\emph{Women are participating, but their performance does not get the same positive reaction as men's}''. Gender bias also appears in OSS communities through the language used. P10 (see TableSM2 in \repo) recommends to ``\emph{\ldots avoid gender pronouns (e.g.: using `guys' is very common, and this gives an idea that it is assumed that contributors are mostly men), so moderating language would help''}. Nafus \cite{nafus2012patches} also mentioned the use of an inadequate language in OSS communities. 

Despite the recognised gender bias, 82.9\% of the respondents report being happy in contributing to OSS and 88.5\%  (strongly) agree that their contributions are well received in open source communities.

We also asked women core developers about the actions that should be taken to create a more inclusive environment for women in OSS. 
To this end, we use an open-ended question and tabulate the answers verbatim (see  TableSM2 in \repo). 
The respondents suggest promoting \emph{women-specific} mentorship programs and events discussing the relevance of contributing to OSS. 
Participants also mentioned more specific initiatives (e.g., Outreachy and local Meetups) that might help to attract more women. 
P9 also emphasizes the relevance of communication, arguing in favor of ``\emph{more women promoting open source projects and work, through blogs, forums, public speaking, (and thus) helping to demystify the
world of open source''}. P23 stressed the importance of increasing confidence as a way of engaging women (cf. recent studies of confidence in context of women in software development~\cite{DBLP:conf/icse/WangWR18,Aivaloglou:Hermans}):
\begin{formal}
\emph{I feel that the solution is to
build confidence.\ldots Every approach towards increasing participation of women has a side that increases
confidence and another that decreases it. The one that I completely support is building a peer group among girls interested in it. It
is not so common to find many girls in technical teams and also, it is not considered cool to be a techie girl. \ldots So, having people around me with whom I can share everything that I do every
day without having a fear of being judged as a freak, has been pretty helpful and encouraging.}
\end{formal}

Participant P20 also recommends events by and for women:
\begin{formal}
\emph{In the R community there are R-Ladies events---held by and for women. Girl only or (predominantly) events in general, could create an environment that suggests that girls and women are actually wanted to be included in the communities}.
\end{formal}
This recommendation agrees with the work of Singh~\cite{DBLP:conf/icse/Singh19} suggesting that women-only spaces in OSS foster discussions, support and empowerment of minorities.

Finally, P2 pointed out that the communities should ``\emph{Stop treating women developers as `women developers' and start treating them as developers}''. This quote suggests both presence of gender bias and frustration caused by it. This can also be seen as a call to support code of conducts~\cite{DBLP:conf/wcre/TouraniAS17} recommending developers to avoid any behavior that might be understood as non-inclusive. This call has been reiterated by P6: ``\emph{OSS Communities should be inclusive not only for women, but for all (men, women, LGBT..., disabled, etc).}''.
Other recommendations also include promoting more women to senior roles and ``\emph{let know the woman not to fear when contributing, she will be treated just like anyone else: good contribution then its accepted regardless of gender, political views or religion, country of origin etc}''.  

\section{Discussion}
\label{sec:discussion}
Below we summarise the insights obtained from our study.

\subsection{Observations}
\paragraph{Vertical segregation in OSS communities.}
Although the underrepresentation of women in OSS communities has been studied before, we found that even less women are core developers in OSS projects. 
This clearly indicates a non-inclusive situation, where women do not appear in top positions, a phenomenon known as \emph{vertical sex segregation}~\cite{benschop2006small}. 
Some respondents suggest that the underrepresentation of women in OSS communities might be due to the low number of women attending undergraduate courses in Computer Science and related subjects. Nonetheless, the percentage of women attending these courses tends to be  10--20\%~\cite{DBLP:journals/te/Olmedo-TorreCBL18, christie2017understanding, DBLP:conf/sigcse/RheingansDDI18, DBLP:conf/icse/Borsotti18}, while the participation of women in OSS is ca. 5\%~\cite{DBLP:conf/icse/LeeC19, StackOverflow,DBLP:conf/chi/VasilescuPRBSDF15}.
We observed that the participation of women as core developers, according to the Truck Factor, is even lower, ca. 2\%. Since participation in OSS communities might help developers to find new job opportunities, it is important to also encourage women to contribute to OSS initiatives. 

\paragraph{There are no (statistically significant) gender-related differences in the work practices of core developers.} 
Men and women core developers similarly 
contribute to the development of OSS. We have also observed women performing managerial activities in some projects: as future work we consider identifying project characteristics that might encourage the participation of women in top positions. 

\paragraph{Perception of gender biases by Women core developers.}
One third of the women core developers faced gender bias at least once. 
This finding stresses importance of social, as opposed to technical, barriers to the participation of women in OSS~\cite{DBLP:journals/cscw/SteinmacherGCR19}.
Moreover, we found that 88.50\% of core developers stated that their contributions are well received by the open source community regardless of their gender and 65.7\% stated that they never had any problems with their contributions and never suffered gender bias. The perception of core developers is different from perception of women developers reported in the  previous research~\cite{DBLP:conf/icse/LeeC19, nafus2012patches, DBLP:conf/icse/ImtiazMCRBM19, DBLP:conf/icse/WangWR19, DBLP:journals/peerj-cs/TerrellKMRMPS17}. 
These high percentages might be expected: women not feeling happy contributing to OSS or not feeling appreciated can be expected to be more likely to leave OSS before becoming core developers.
At the same time this difference calls for a more careful investigation of the differences between core developers and non-core developers, and differences in perception of their contributions.

\subsection{Suggestions to make OSS more inclusive}
Survey respondents made several suggestions how to make OSS more inclusive for women.
We hope that these suggestions might make OSS more welcoming other  minorities as well (cf.~\cite{DBLP:conf/icse/FordMS19}):
\begin{enumerate}
\item Promote women-specific mentorship programs, akin to discussed by Hyrynsalmi~\cite{DBLP:conf/icse/Hyrynsalmi19} and Buhnova and Prikrylova~\cite{DBLP:conf/icse/BuhnovaP19}. 
\item Promote women to senior roles: some communities, e.g., Open Stack, already implement this~\cite{DBLP:journals/software/IzquierdoHSR19}. 
\item Organize women-specific events, such as local meetups, or even tech groups~\cite{DBLP:conf/icse/Singh19}. Such events are organised, e.g., by R ladies group or the Pyladies group. 

\item Avoid gendered language (e.g., using `guys' when a `folks' would work). 
\end{enumerate}
The four suggestions can be combined and should contribute to women's confidence to contribute to OSS communities~\cite{DBLP:conf/icse/SilveiraMMVP19,DBLP:conf/icse/WangWR18}.

\section{Threats to Validity}\label{sec:threats}

As any empirical work, this work also has many limitations and threats to validity.

{\bf Construct Validity.} 
The main construct used in this study is the construct of `gender''. Gender is a complex social construct and no automatic tool can capture its entire complexity. Moreover, the accuracy of gender classifiers is inherently limited by the information developers provide in the software repositories. Many users do not use their real names, so we might not reliably extracted gender information~\cite{DBLP:conf/icse/QiuNBSV19}. To minimize this threat we (a) combine the results of two independently developed gender classifiers, and (b) manually validate the gender of women core developers, by checking information publicly available on social networks. We were able to validate the gender from 39 (out of 45) women core developers---without discarding the same women core developers contributing to more than one project. However, manual identification of gender might introduce bias as researchers necessarily can only indicate gender as perceived by the outsiders based on gender expression rather than gender identity.
Another construct we use in this study is the construct of a ``core developer''. To this end we use the notion of a Truck Factor, and specifically the implementation of the Truck Factor detection proposed by Avelino et al.~\cite{DBLP:conf/iwpc/AvelinoPHV16}.
While the approach of Avelino et al. has outperform the competing techniques in the evaluation study of Ferreira et al.~\cite{DBLP:conf/iwpc/FerreiraVF17}, and hence can be seen as state-of-the-art, as any automatic approach it can never be expected to perfectly identify developers whose departure from the project will heavily affect the project maintenance.

\vspace{0.2cm}
\noindent
{\bf Internal Validity.}
The Truck Factor identification approach we used introduces additional threats. In particular, it only outputs the core developer's name. For this reason, we have to search the name of the contributor in the GitHub API to identify the corresponding user id. We accept the first value returned in the search performed, which can be a threat since we have no way of guaranteeing that the first value returned is, in fact, the login corresponding to that name. Also, we excluded some names that did not return user ids in the search, meaning that the person probably changed the name, or their account was deleted, or their account privacy setting was changed so that the contributor's name was not displayed. We also excluded names where their content was the same as login.

\vspace{0.2cm}
\noindent
{\bf Conclusion Validity.} We found that 80\% of the women core developers that answered our survey are less than 35 years old and contribute to open source projects for less than five years. This finding brings additional threats to the conclusions of our work. For instance, one might argue that the small number of women core developers contributing to open source projects is due either to lack of experience or lack of interest to continue contributing to the development of open-source systems for long periods. 

\vspace{0.2cm}
\noindent
{\bf External Validity.} We did not consider all possible open source projects available out there. Although we covered hundreds of projects hosted on GitHub, many other open source projects are hosted on different forges (e.g., GitLab or BitBucket). However, we do not expect major changes in our results, since we consider our projects' population diverse enough (in terms of programming languages used, number of core members, etc.). The survey was answered by 35 developers. This low number is due to the low percentage of women core developers in projects. Therefore, the representativeness of the sample is high, considering that among almost 2000 core developers, women correspond to a really small fraction, only 2.3\% of the core developers are women. Thus, we can consider that the response rate was high (35 responses) out of a total of 39 women.

\section{Conclusion}
\label{remark}

In this work we studied a different interpretation of gender-bias in open-source communities, i.e., \emph{vertical sex segregation}~\cite{benschop2006small}, which occurs when the participation of men and women in top positions is even less balanced. In our study, ``top positions'' correspond to the core developers of a system, which we identified using the Truck Factor. 
We found a more significant underrepresentation of
women core developers than women developers---only 45 in \projects open-source systems have at least one woman core developer, and only 2.30\% of the total number of core developers are women.
Nonetheless, this group of women core developers contribute in a similar fashion as other groups of men key developers (considering frequency and size of contributing). 

Women core developers believe that gender diversity is important for OSS communities. According to their opinions, gender diversity might contribute to improving the communication among team members and help to generate different ideas while designing a software products. Nonetheless, 34.7\% of the women core developers we surveyed in our study report having faced some sort of gender bias (11.4\% of them claim often facing gender bias). This group also consider that promoting women specific events can contribute to making open-source communities more inclusive.

\section*{Acknowledgements}

This research work has the support of the Research Support Foundation of the Federal District (FAPDF), research grant 05/2018 and CNPq (grant 309032/2019-9).

\balance

\bibliography{references}

\begin{thebibliography}{58}
\providecommand{\natexlab}[1]{#1}
\providecommand{\url}[1]{\texttt{#1}}
\expandafter\ifx\csname urlstyle\endcsname\relax
  \providecommand{\doi}[1]{doi: #1}\else
  \providecommand{\doi}{doi: \begingroup \urlstyle{rm}\Url}\fi

\bibitem[Vasilescu et~al.(2015)Vasilescu, Posnett, Ray, van~den Brand,
  Serebrenik, Devanbu, and Filkov]{DBLP:conf/chi/VasilescuPRBSDF15}
Bogdan Vasilescu, Daryl Posnett, Baishakhi Ray, Mark G.~J. van~den Brand,
  Alexander Serebrenik, Premkumar~T. Devanbu, and Vladimir Filkov.
\newblock Gender and tenure diversity in github teams.
\newblock In \emph{{CHI}}, pages 3789--3798, 10.1145/2702123.2702549, 2015.
  {ACM}.

\bibitem[Catolino et~al.(2019)Catolino, Palomba, Tamburri, Serebrenik, and
  Ferrucci]{Catolino:2019:GDW:3339974.3339977}
Gemma Catolino, Fabio Palomba, Damian~A. Tamburri, Alexander Serebrenik, and
  Filomena Ferrucci.
\newblock Gender diversity and women in software teams: How do they affect
  community smells?
\newblock In \emph{ICSE-SEIS}, pages 11--20, 10.1109/ICSE-SEIS.2019.00010,
  2019. IEEE.

\bibitem[Hui and Farnham(2016)]{DBLP:conf/group/HuiF16}
Julie~S. Hui and Shelly~Diane Farnham.
\newblock Designing for inclusion: Supporting gender diversity in independent
  innovation teams.
\newblock In \emph{{GROUP}}, pages 71--85, ACM, 2016. {ACM}.

\bibitem[Brown and Parker(2019)]{google2019}
Danielle Brown and Melonie Parker.
\newblock Google diversity annual report 2019, 2019.
\newblock URL \url{https://diversity.google/}.
\newblock (Date last accessed 16-April-2019).

\bibitem[Imtiaz et~al.(2019)Imtiaz, Middleton, Chakraborty, Robson, Bai, and
  Murphy{-}Hill]{DBLP:conf/icse/ImtiazMCRBM19}
Nasif Imtiaz, Justin Middleton, Joymallya Chakraborty, Neill Robson, Gina Bai,
  and Emerson~R. Murphy{-}Hill.
\newblock Investigating the effects of gender bias on github.
\newblock In \emph{{ICSE}}, pages 700--711, 10.1109/ICSE.2019.00079, 2019.
  {IEEE} / {ACM}.

\bibitem[Wang and Redmiles(2019)]{DBLP:conf/icse/WangWR19}
Yi~Wang and David Redmiles.
\newblock Implicit gender biases in professional software development: An
  empirical study.
\newblock In \emph{{ICSE - Software Engineering in Society}}, pages 1--10,
  https://bit.ly/2Hu42MJ, 2019. {ACM}.

\bibitem[Daniel~Izquierdo and Price(2017)]{OpenStack2017}
Nicole~Huesman Daniel~Izquierdo and Allison Price.
\newblock Gender diversity in the openstack community: A new report, 2017.
\newblock URL
  \url{https://superuser.openstack.org/wp-content/uploads/2017/07/OpenStack-Gender-Diversity-Report\_Apr2017.pdf}.
\newblock (Date last accessed 16-April-2019).

\bibitem[Izquierdo et~al.(2019)Izquierdo, Huesman, Serebrenik, and
  Robles]{DBLP:journals/software/IzquierdoHSR19}
Daniel Izquierdo, Nicole Huesman, Alexander Serebrenik, and Gregorio Robles.
\newblock Openstack gender diversity report.
\newblock \emph{{IEEE} Software}, 36\penalty0 (1):\penalty0 28--33, 2019.

\bibitem[Robles et~al.(2014)Robles, {Arjona Reina}, Serebrenik, Vasilescu, and
  Gonz{\'{a}}lez{-}Barahona]{DBLP:conf/msr/RoblesRSVG14}
Gregorio Robles, Laura {Arjona Reina}, Alexander Serebrenik, Bogdan Vasilescu,
  and Jes{\'{u}}s~M. Gonz{\'{a}}lez{-}Barahona.
\newblock {FLOSS} 2013: a survey dataset about free software contributors:
  challenges for curating, sharing, and combining.
\newblock In \emph{{MSR}}, pages 396--399, 10.1145/2597073.2597129, 2014.
  {ACM}.

\bibitem[Lin and Serebrenik(2016)]{DBLP:conf/msr/0008S16}
Bin Lin and Alexander Serebrenik.
\newblock Recognizing gender of stack overflow users.
\newblock In \emph{{MSR}}, pages 425--429, 0.1145/2901739.2901777, 2016. {ACM}.

\bibitem[Ortu et~al.(2017)Ortu, Destefanis, Counsell, Swift, Tonelli, and
  Marchesi]{DBLP:journals/jserd/OrtuDCSTM17}
Marco Ortu, Giuseppe Destefanis, Steve Counsell, Stephen Swift, Roberto
  Tonelli, and Michele Marchesi.
\newblock How diverse is your team? investigating gender and nationality
  diversity in github teams.
\newblock \emph{J. Software Eng. R{\&}D}, 5:\penalty0 9, 2017.

\bibitem[Gila et~al.(2014)Gila, Jaafa, Omar, and Tunio]{gila2014impact}
Abdul~Rehman Gila, Jafreezal Jaafa, Mazni Omar, and Muhammad~Zahid Tunio.
\newblock Impact of personality and gender diversity on software development
  teams' performance.
\newblock In \emph{2014 International Conference on Computer, Communications,
  and Control Technology (I4CT)}, pages 261--265, IEEE, 2014. IEEE.

\bibitem[Nafus(2012)]{nafus2012patches}
Dawn Nafus.
\newblock ‘patches don’t have gender’: What is not open in open source
  software.
\newblock \emph{New Media \& Society}, 14\penalty0 (4):\penalty0 669--683,
  2012.

\bibitem[Garc{\'i}a-Gonz{\'a}lez et~al.(2019)Garc{\'i}a-Gonz{\'a}lez,
  Forc{\'e}n, and Jimenez-Sanchez]{10.1371/journal.pone.0225763}
Judit Garc{\'i}a-Gonz{\'a}lez, Patricia Forc{\'e}n, and Maria Jimenez-Sanchez.
\newblock Men and women differ in their perception of gender bias in research
  institutions.
\newblock \emph{PLOS ONE}, 14\penalty0 (12):\penalty0 1--21, 12 2019.
\newblock \doi{10.1371/journal.pone.0225763}.
\newblock URL \url{https://doi.org/10.1371/journal.pone.0225763}.

\bibitem[Derks et~al.(2011)Derks, Laar, Ellemers, and
  de~Groot]{doi:10.1177/0956797611417258}
Belle Derks, Colette~Van Laar, Naomi Ellemers, and Kim de~Groot.
\newblock Gender-bias primes elicit queen-bee responses among senior
  policewomen.
\newblock \emph{Psychological Science}, 22\penalty0 (10):\penalty0 1243--1249,
  2011.
\newblock \doi{10.1177/0956797611417258}.
\newblock URL \url{https://doi.org/10.1177/0956797611417258}.
\newblock PMID: 21873568.

\bibitem[Benschop(2006)]{benschop2006small}
Yvonne Benschop.
\newblock Of small steps and the longing for giant leaps.
\newblock \emph{Handbook of workplace diversity}, 1:\penalty0 274--298, 2006.

\bibitem[Campos-Soria et~al.(2011)Campos-Soria, Marchante-Mera, and
  Ropero-Garc{\'\i}a]{campos2011patterns}
Juan~Antonio Campos-Soria, Andr{\'e}s Marchante-Mera, and Miguel~Angel
  Ropero-Garc{\'\i}a.
\newblock Patterns of occupational segregation by gender in the hospitality
  industry.
\newblock \emph{International Journal of Hospitality Management}, 30\penalty0
  (1):\penalty0 91--102, 2011.

\bibitem[Ricca et~al.(2011)Ricca, Marchetto, and
  Torchiano]{DBLP:conf/profes/RiccaMT11}
Filippo Ricca, Alessandro Marchetto, and Marco Torchiano.
\newblock On the difficulty of computing the truck factor.
\newblock In \emph{{PROFES}}, volume 6759 of \emph{Lecture Notes in Business
  Information Processing}, pages 337--351,
  https://doi.org/10.1007/978-3-642-21843-9\_26, 2011. Springer.

\bibitem[Cosentino et~al.(2015)Cosentino, Izquierdo, and
  Cabot]{DBLP:conf/wcre/CosentinoIC15}
Valerio Cosentino, Javier Luis~C{\'{a}}novas Izquierdo, and Jordi Cabot.
\newblock Assessing the bus factor of git repositories.
\newblock In \emph{{SANER}}, pages 499--503, 10.1109/SANER.2015.7081864, 2015.
  {IEEE} Computer Society.

\bibitem[Avelino et~al.(2016)Avelino, Passos, Hora, and
  Valente]{DBLP:conf/iwpc/AvelinoPHV16}
Guilherme Avelino, Leonardo~Teixeira Passos, Andr{\'{e}}~C. Hora, and
  Marco~Tulio Valente.
\newblock A novel approach for estimating truck factors.
\newblock In \emph{{ICPC}}, pages 1--10, IEEE, 2016. {IEEE} Computer Society.

\bibitem[Vasilescu et~al.(2014)Vasilescu, Capiluppi, and
  Serebrenik]{DBLP:journals/iwc/VasilescuCS14}
Bogdan Vasilescu, Andrea Capiluppi, and Alexander Serebrenik.
\newblock Gender, representation and online participation: {A} quantitative
  study.
\newblock \emph{Interacting with Computers}, 26\penalty0 (5):\penalty0
  488--511, 2014.

\bibitem[Borsotti(2018)]{DBLP:conf/icse/Borsotti18}
Valeria Borsotti.
\newblock Barriers to gender diversity in software development education:
  actionable insights from a danish case study.
\newblock In \emph{{ICSE} {(SEET)}}, pages 146--152, {ACM}, 2018. {ACM}.

\bibitem[Bhattacharya et~al.(2018)Bhattacharya, Bhattacharya, and
  Mohapatra]{DBLP:journals/ijhcitp/BhattacharyaBM18}
Shubhasheesh Bhattacharya, Sonali Bhattacharya, and Sweta Mohapatra.
\newblock Enablers for advancement of women into leadership position: {A} study
  based on {IT/ITES} sector in india.
\newblock \emph{{IJHCITP}}, 9\penalty0 (4):\penalty0 1--22, 2018.

\bibitem[Maciel et~al.(2018)Maciel, Bim, and
  da~Silva~Figueiredo]{DBLP:conf/icse/MacielBF18}
Cristiano Maciel, Silvia~Am{\'{e}}lia Bim, and Karen da~Silva~Figueiredo.
\newblock Digital girls program - disseminating computer science to girls in
  brazil.
\newblock In \emph{GE@ICSE}, pages 29--32,
  https://ieeexplore.ieee.org/document/8452748, 2018. {ACM}.

\bibitem[Botturi et~al.(2012)Botturi, Bramani, and
  McCusker]{DBLP:journals/jucs/BotturiBM12}
Luca Botturi, Chiara Bramani, and Sean McCusker.
\newblock Boys are like girls: Insights in the gender digital divide in higher
  education in switzerland and europe.
\newblock \emph{J. {UCS}}, 18\penalty0 (3):\penalty0 353--376, 2012.

\bibitem[Rheingans et~al.(2018)Rheingans, D'Eramo, Diaz{-}Espinoza, and
  Ireland]{DBLP:conf/sigcse/RheingansDDI18}
Penny Rheingans, Erica D'Eramo, Crystal Diaz{-}Espinoza, and Danyelle Ireland.
\newblock A model for increasing gender diversity in technology.
\newblock In \emph{{SIGCSE}}, pages 459--464, 10.1145/3159450.3159533, 2018.
  {ACM}.

\bibitem[de~Ribaupierre et~al.(2018)de~Ribaupierre, Jones, Loizides, and
  Cherdantseva]{DBLP:conf/icse/RibaupierreJLC18}
H{\'{e}}l{\`{e}}ne de~Ribaupierre, Kathryn Jones, Fernando Loizides, and Yulia
  Cherdantseva.
\newblock Towards gender equality in software engineering: The {NSA} approach.
\newblock In \emph{GE@ICSE}, pages 10--13, {ACM}, 2018. {ACM}.

\bibitem[Jr. et~al.(2005)Jr., Schulte, and
  Giguette]{DBLP:conf/sigcse/LopezSG05}
Antonio M.~Lopez Jr., Lisa~J. Schulte, and Marguerite~S. Giguette.
\newblock Climbing onto the shoulders of giants.
\newblock In \emph{{SIGCSE}}, pages 401--405, 10.1145/1047344.1047477, 2005.
  {ACM}.

\bibitem[Botella et~al.(2019)Botella, Rueda, L{\'{o}}pez{-}I{\~{n}}esta, and
  Marzal]{DBLP:journals/entropy/BotellaRLM19}
Carmen Botella, Silvia Rueda, Emilia L{\'{o}}pez{-}I{\~{n}}esta, and Paula
  Marzal.
\newblock Gender diversity in {STEM} disciplines: {A} multiple factor problem.
\newblock \emph{Entropy}, 21\penalty0 (1):\penalty0 30, 2019.

\bibitem[Bosu and Sultana(2019)]{DBLP:conf/esem/BosuS19}
Amiangshu Bosu and Kazi~Zakia Sultana.
\newblock Diversity and inclusion in open source software {(OSS)} projects:
  Where do we stand?
\newblock In \emph{{ESEM}}, pages 1--11, IEEE, 2019. {ACM/IEEE International
  Symposium on Empirical Software Engineering and Measurement (ESEM)}.
\newblock \doi{10.1109/ESEM.2019.8870179}.

\bibitem[Catolino et~al.(2020)Catolino, Palomba, Tamburri, Serebrenik, and
  Ferrucci]{DBLP:journals/software/CatolinoPTSF20}
Gemma Catolino, Fabio Palomba, Damian~A. Tamburri, Alexander Serebrenik, and
  Filomena Ferrucci.
\newblock Gender diversity and community smells: Insights from the trenches.
\newblock \emph{{IEEE} Software}, 37\penalty0 (1):\penalty0 10--16, 2020.
\newblock \doi{10.1109/MS.2019.2944594}.
\newblock URL \url{https://doi.org/10.1109/MS.2019.2944594}.

\bibitem[Qiu et~al.(2019)Qiu, Nolte, Brown, Serebrenik, and
  Vasilescu]{DBLP:conf/icse/QiuNBSV19}
Huilian~Sophie Qiu, Alexander Nolte, Anita Brown, Alexander Serebrenik, and
  Bogdan Vasilescu.
\newblock Going farther together: the impact of social capital on sustained
  participation in open source.
\newblock In \emph{{ICSE}}, pages 688--699, 10.1109/ICSE.2019.00078, 2019.
  {IEEE} / {ACM}.

\bibitem[Balali et~al.(2018)Balali, Steinmacher, Annamalai, Sarma, and
  Gerosa]{DBLP:journals/cscw/BalaliSASG18}
Sogol Balali, Igor Steinmacher, Umayal Annamalai, Anita Sarma, and
  Marco~Aur{\'{e}}lio Gerosa.
\newblock Newcomers' barriers. . . is that all? an analysis of mentors' and
  newcomers' barriers in {OSS} projects.
\newblock \emph{Computer Supported Cooperative Work}, 27\penalty0
  (3-6):\penalty0 679--714, 2018.

\bibitem[Lee and Carver(2019)]{DBLP:conf/icse/LeeC19}
Amanda Lee and Jeffrey~C. Carver.
\newblock {FLOSS} participants' perceptions about gender and inclusiveness: a
  survey.
\newblock In \emph{{ICSE}}, pages 677--687, 10.1109/ICSE.2019.00077, 2019.
  {IEEE} / {ACM}.

\bibitem[Baltes and Ralph(2020)]{baltes2020sampling}
Sebastian Baltes and Paul Ralph.
\newblock Sampling in software engineering research: A critical review and
  guidelines.
\newblock \emph{arXiv preprint arXiv:2002.07764}, 1:\penalty0 1--21, 2020.

\bibitem[Pinto et~al.(2016)Pinto, Steinmacher, and Gerosa]{casual-developers}
Gustavo Pinto, Igor Steinmacher, and Marco~Aur{\'{e}}lio Gerosa.
\newblock More common than you think: An in-depth study of casual contributors.
\newblock In \emph{{IEEE} 23rd International Conference on Software Analysis,
  Evolution, and Reengineering, {SANER} 2016, Suita, Osaka, Japan, March 14-18,
  2016 - Volume 1}, pages 112--123, https://doi.org/10.1109/SANER.2016.68,
  2016. {IEEE} Computer Society.
\newblock \doi{10.1109/SANER.2016.68}.

\bibitem[Borges et~al.(2016)Borges, Hora, and Valente]{gh-popularity}
Hudson Borges, Andre Hora, and Marco~Tulio Valente.
\newblock Predicting the popularity of github repositories.
\newblock In \emph{Proceedings of the The 12th International Conference on
  Predictive Models and Data Analytics in Software Engineering}, PROMISE 2016,
  New York, NY, USA, 2016. Association for Computing Machinery.
\newblock ISBN 9781450347723.
\newblock \doi{10.1145/2972958.2972966}.
\newblock URL \url{https://doi.org/10.1145/2972958.2972966}.

\bibitem[Ferreira et~al.(2017)Ferreira, Valente, and
  Ferreira]{DBLP:conf/iwpc/FerreiraVF17}
M{\'{\i}}vian~M. Ferreira, Marco~Tulio Valente, and Kecia Aline~M. Ferreira.
\newblock A comparison of three algorithms for computing truck factors.
\newblock In \emph{{ICPC}}, pages 207--217, IEEE, 2017. {IEEE} Computer
  Society.

\bibitem[Hattori and Lanza(2008)]{hattori:ase-2008}
Lile Hattori and Michele Lanza.
\newblock On the nature of commits.
\newblock In \emph{23rd {IEEE/ACM} International Conference on Automated
  Software Engineering - Workshop Proceedings {(ASE} Workshops 2008), 15-16
  September 2008, L'Aquila, Italy}, pages 63--71, dblp computer science
  bibliography, https://dblp.org, 2008. {IEEE}.
\newblock \doi{10.1109/ASEW.2008.4686322}.
\newblock URL \url{https://doi.org/10.1109/ASEW.2008.4686322}.

\bibitem[Maindonald and Braun(2010)]{data-analysis-using-R}
John Maindonald and W.~John Braun.
\newblock \emph{Data Analysis and Graphics Using R: An Example-Based Approach}.
\newblock Cambridge University Press, USA, 3rd edition, 2010.
\newblock ISBN 0521762936.

\bibitem[Konietschke et~al.(2015)Konietschke, Placzek, Schaarschmidt, and
  Hothorn]{JSSv064i09}
Frank Konietschke, Marius Placzek, Frank Schaarschmidt, and Ludwig Hothorn.
\newblock nparcomp: An r software package for nonparametric multiple
  comparisons and simultaneous confidence intervals.
\newblock \emph{Journal of Statistical Software, Articles}, 64\penalty0
  (9):\penalty0 1--17, 2015.
\newblock ISSN 1548-7660.
\newblock \doi{10.18637/jss.v064.i09}.
\newblock URL \url{https://www.jstatsoft.org/v064/i09}.

\bibitem[Vasilescu et~al.(2013)Vasilescu, Filkov, and
  Serebrenik]{vasilescu:social2013}
Bogdan Vasilescu, Vladimir Filkov, and Alexander Serebrenik.
\newblock Stackoverflow and github: Associations between software development
  and crowdsourced knowledge.
\newblock In \emph{International Conference on Social Computing, SocialCom
  2013, SocialCom/PASSAT/BigData/EconCom/BioMedCom 2013, Washington, DC, USA,
  8-14 September, 2013}, pages 188--195,
  https://doi.org/10.1109/SocialCom.2013.35, 2013. {IEEE} Computer Society.
\newblock \doi{10.1109/SocialCom.2013.35}.

\bibitem[Dunnett(1955)]{dunnett1955multiple}
Charles~W Dunnett.
\newblock A multiple comparison procedure for comparing several treatments with
  a control.
\newblock \emph{Journal of the American Statistical Association}, 50\penalty0
  (272):\penalty0 1096--1121, 1955.

\bibitem[Kitchenham and Pfleeger(2008)]{DBLP:books/sp/08/KitchenhamP08}
Barbara~A. Kitchenham and Shari~Lawrence Pfleeger.
\newblock Personal opinion surveys.
\newblock In \emph{Guide to Advanced Empirical Software Engineering}, pages
  63--92. Springer, 10.1007/978-1-84800-044-5\_3, 2008.

\bibitem[Avelino et~al.(2019)Avelino, Constantinou, Valente, and
  Serebrenik]{DBLP:conf/esem/AvelinoCVS19}
Guilherme Avelino, Eleni Constantinou, Marco~Tulio Valente, and Alexander
  Serebrenik.
\newblock On the abandonment and survival of open source projects: An empirical
  investigation.
\newblock In \emph{{ESEM}}, pages 1--12, 10.1109/ESEM.2019.8870181, 2019.
  {IEEE}.

\bibitem[Wang et~al.(2018)Wang, Wang, and Redmiles]{DBLP:conf/icse/WangWR18}
Zhendong Wang, Yi~Wang, and David~F. Redmiles.
\newblock Competence-confidence gap: a threat to female developers'
  contribution on github.
\newblock In Val{\'{e}}rie Issarny and Schahram Dustdar, editors,
  \emph{Proceedings of the 40th International Conference on Software
  Engineering: Software Engineering in Society, {ICSE} {(SEIS)} 2018,
  Gothenburg, Sweden, May 27 - June 03, 2018}, pages 81--90,
  https://doi.org/10.1145/3183428.3183437, 2018. {ACM}.
\newblock \doi{10.1145/3183428.3183437}.

\bibitem[Aivaloglou and Hermans(2019)]{Aivaloglou:Hermans}
Efthimia Aivaloglou and Felienne Hermans.
\newblock How is programming taught in code clubs? exploring the experiences
  and gender perceptions of code club teachers.
\newblock In \emph{Proceedings of the 19th Koli Calling International
  Conference on Computing Education Research}, Koli Calling ’19, New York,
  NY, USA, 2019. Association for Computing Machinery.
\newblock ISBN 9781450377157.
\newblock \doi{10.1145/3364510.3364514}.
\newblock URL \url{https://doi.org/10.1145/3364510.3364514}.

\bibitem[Singh(2019)]{DBLP:conf/icse/Singh19}
Vandana Singh.
\newblock Women-only spaces of open source.
\newblock In Ivica Crnkovic, Karina~Kohl Silveira, and Sara Sprenkle, editors,
  \emph{Proceedings of the 2nd International Workshop on Gender Equality in
  Software Engineering, GE@ICSE 2019, Montreal, QC, Canada, May 27, 2019},
  pages 17--20, https://doi.org/10.1109/GE.2019.00010, 2019. {IEEE} / {ACM}.
\newblock \doi{10.1109/GE.2019.00010}.

\bibitem[Tourani et~al.(2017)Tourani, Adams, and
  Serebrenik]{DBLP:conf/wcre/TouraniAS17}
Parastou Tourani, Bram Adams, and Alexander Serebrenik.
\newblock Code of conduct in open source projects.
\newblock In Martin Pinzger, Gabriele Bavota, and Andrian Marcus, editors,
  \emph{{IEEE} 24th International Conference on Software Analysis, Evolution
  and Reengineering, {SANER} 2017, Klagenfurt, Austria, February 20-24, 2017},
  pages 24--33, https://doi.org/10.1109/SANER.2017.7884606, 2017. {IEEE}
  Computer Society.
\newblock \doi{10.1109/SANER.2017.7884606}.

\bibitem[Olmedo{-}Torre et~al.(2018)Olmedo{-}Torre, Carracedo, Ballesteros,
  L{\'{o}}pez, Perez{-}Poch, and
  Lopez{-}Beltran]{DBLP:journals/te/Olmedo-TorreCBL18}
Noelia Olmedo{-}Torre, Ferm{\'{\i}}n~S{\'{a}}nchez Carracedo, M.~Nuria~Salan
  Ballesteros, David L{\'{o}}pez, Antoni Perez{-}Poch, and Mireia
  Lopez{-}Beltran.
\newblock Do female motives for enrolling vary according to {STEM} profile?
\newblock \emph{{IEEE} Trans. Education}, 61\penalty0 (4):\penalty0 289--297,
  2018.

\bibitem[Christie et~al.(2017)Christie, O’Neill, Rutter, Young, and
  Medland]{christie2017understanding}
Michael Christie, Maureen O’Neill, Kerry Rutter, Graham Young, and Angeline
  Medland.
\newblock Understanding why women are under-represented in science, technology,
  engineering and mathematics (stem) within higher education: A regional case
  study.
\newblock \emph{Production}, 27\penalty0 (SPE):\penalty0 1--9, 2017.

\bibitem[Overflow(2019)]{StackOverflow}
Stack Overflow.
\newblock Developer survey results 2019, 2019.
\newblock URL \url{https://insights.stackoverflow.com/survey/2019}.
\newblock (Date last accessed 18-April-2019).

\bibitem[Steinmacher et~al.(2019)Steinmacher, Gerosa, Conte, and
  Redmiles]{DBLP:journals/cscw/SteinmacherGCR19}
Igor Steinmacher, Marco~Aur{\'{e}}lio Gerosa, Tayana~Uch{\^{o}}a Conte, and
  David~F. Redmiles.
\newblock Overcoming social barriers when contributing to open source software
  projects.
\newblock \emph{Computer Supported Cooperative Work}, 28\penalty0
  (1-2):\penalty0 247--290, 2019.

\bibitem[Terrell et~al.(2017)Terrell, Kofink, Middleton, Rainear,
  Murphy{-}Hill, Parnin, and Stallings]{DBLP:journals/peerj-cs/TerrellKMRMPS17}
Josh Terrell, Andrew Kofink, Justin Middleton, Clarissa Rainear, Emerson~R.
  Murphy{-}Hill, Chris Parnin, and Jon Stallings.
\newblock Gender differences and bias in open source: pull request acceptance
  of women versus men.
\newblock \emph{PeerJ Computer Science}, 3:\penalty0 e111, 2017.

\bibitem[Ford et~al.(2019)Ford, Milewicz, and
  Serebrenik]{DBLP:conf/icse/FordMS19}
Denae Ford, Reed Milewicz, and Alexander Serebrenik.
\newblock How remote work can foster a more inclusive environment for
  transgender developers.
\newblock In Ivica Crnkovic, Karina~Kohl Silveira, and Sara Sprenkle, editors,
  \emph{Proceedings of the 2nd International Workshop on Gender Equality in
  Software Engineering, GE@ICSE 2019, Montreal, QC, Canada, May 27, 2019},
  pages 9--12, https://doi.org/10.1109/GE.2019.00011, 2019. {IEEE} / {ACM}.
\newblock \doi{10.1109/GE.2019.00011}.

\bibitem[Hyrynsalmi(2019)]{DBLP:conf/icse/Hyrynsalmi19}
Sonja~M. Hyrynsalmi.
\newblock The underrepresentation of women in the software industry: thoughts
  from career-changing women.
\newblock In Ivica Crnkovic, Karina~Kohl Silveira, and Sara Sprenkle, editors,
  \emph{Proceedings of the 2nd International Workshop on Gender Equality in
  Software Engineering, GE@ICSE 2019, Montreal, QC, Canada, May 27, 2019},
  pages 1--4, https://doi.org/10.1109/GE.2019.00008, 2019. {IEEE} / {ACM}.
\newblock \doi{10.1109/GE.2019.00008}.

\bibitem[Buhnova and Prikrylova(2019)]{DBLP:conf/icse/BuhnovaP19}
Barbora Buhnova and Dita Prikrylova.
\newblock Women want to learn tech: lessons from the czechitas education
  project.
\newblock In Ivica Crnkovic, Karina~Kohl Silveira, and Sara Sprenkle, editors,
  \emph{Proceedings of the 2nd International Workshop on Gender Equality in
  Software Engineering, GE@ICSE 2019, Montreal, QC, Canada, May 27, 2019},
  pages 25--28, https://doi.org/10.1109/GE.2019.00013, 2019. {IEEE} / {ACM}.
\newblock \doi{10.1109/GE.2019.00013}.

\bibitem[Silveira et~al.(2019)Silveira, Musse, Manssour, Vieira, and
  Prikladnicki]{DBLP:conf/icse/SilveiraMMVP19}
Karina~Kohl Silveira, Soraia~Raupp Musse, Isabel~Harb Manssour, Renata Vieira,
  and Rafael Prikladnicki.
\newblock Confidence in programming skills: gender insights from stackoverflow
  developers survey.
\newblock In \emph{{ICSE} (Companion Volume)}, pages 234--235,
  10.1109/ICSE-Companion.2019.00091, 2019. {IEEE} / {ACM}.

\end{thebibliography}

\end{document}